# AMRScan 2: provenance-recorded detection of antimicrobial resistance determinants in bacterial assemblies, with archived susceptibility data from NCBI Pathogen Detection

Kaitao Lai[1,2]

[1] Faculty of Medicine and Health, The University of Sydney, NSW 2006, Australia; [2] ANZAC Research Institute, Hospital Road, Concord NSW 2137, Australia

Correspondence: kaitao.lai@sydney.edu.au

## Abstract

Studies that relate bacterial genomes to antimicrobial susceptibility test (AST) results need determinant calls and phenotypes that can be traced to the exact genome, software, database and source export they came from. AMRScan 2 is a rewrite of the AMRScan toolkit for this purpose. Its Nextflow workflow runs NCBI AMRFinderPlus on genome assemblies in a pinned container, writes a provenance record for every sample with the command, the software version and the SHA-256 hashes of the input, of every database file and of the report, and converts each report into a table with a fixed schema that keeps every original column. The conversion step stops if a report no longer matches its provenance record. A second tool archives AST and isolate tables exported from NCBI Pathogen Detection, checks each export against an independent count, and builds isolate and endpoint tables offline that can be rebuilt and compared byte for byte. AMRScan does not infer phenotype from determinants; calls and phenotypes are kept as separate layers and joined by assembly accession. The caller wrapper was used to call 10,521 public *Escherichia coli* assemblies on a workstation without a failed call, and a run of 1,000 of these assemblies through the workflow on an HPC system gave byte-identical reports. The workflow follows nf-core conventions and is tested with unit tests, nf-test and continuous integration. This version replaces the 2025 description of AMRScan 1, a BLAST search against CARD, which is kept unchanged in the repository.

# Introduction

Genomic detection of antimicrobial resistance (AMR) determinants is now routine, and several tools and workflows run the established detectors at scale. AMRFinderPlus detects acquired genes and, for supported taxa, chromosomal point mutations against NCBI's curated Reference Gene Catalog [1, 2]. abritAMR wraps AMRFinderPlus for accredited use in clinical and public health laboratories [3], staramr reports ResFinder and PointFinder results together with a predicted antibiogram [4], Bactopia runs AMRFinderPlus among many analyses of bacterial genomes [5], nf-core/funcscan runs several AMR detectors on assembled contigs, and hAMRonization converts the outputs of many detectors into one specification [6].

A different need arises when determinant calls are used as data in a study, for example to compare rule-based and learned predictions of AST phenotype across thousands of public genomes. The analysis then depends on knowing, for every isolate, which file was called, with which software build and which database files, and that the table being analysed is the one the caller wrote. Database releases on the distribution site can change after download, intermediate tables can be regenerated with different settings, and public AST records can change between exports. These problems are not specific to any one detector, but they are rarely recorded at the level of the individual sample.

AMRScan 1 [7] was a small toolkit, written once as an R script and once as a single Nextflow file, that searched sequences against the CARD protein homolog model [8] with BLAST [9]. AMRScan 2 is a rewrite with a narrower aim: to produce AMRFinderPlus calls and NCBI Pathogen Detection AST data that can each be traced to their inputs, and to keep the two separate until an analysis joins them. This note describes the design, the checks it makes and its use on a cohort of 10,521 *E. coli* genomes. Version 1 is kept unchanged in the `legacy/` directory of the repository so that its preprint can still be reproduced.

# Implementation

## Workflow

The workflow is written in Nextflow DSL2 [10] and follows nf-core conventions [11]: a parameter schema, samplesheet validation with the nf-schema plugin, meta maps, process labels and a record of tool versions per process (Fig. 1A). It takes assembled nucleotide FASTA files, either as a file pattern or as a samplesheet giving the sample identifier, the path and, optionally, the SHA-256 of each file. Parameters and the samplesheet are checked before any task is scheduled; duplicate or malformed sample identifiers, missing files and a missing database directory stop the run with a message naming the problem.

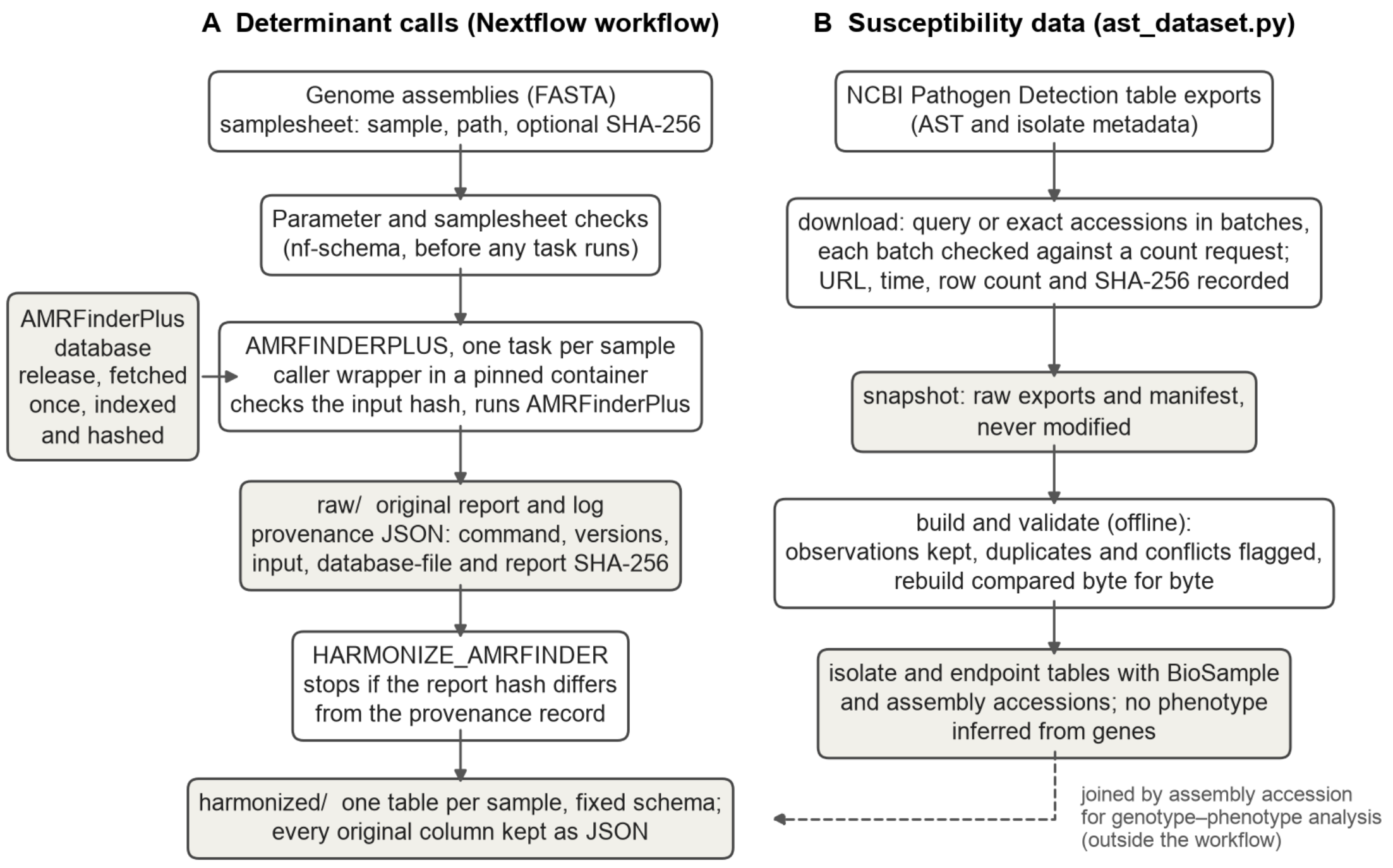

**Fig. 1.** Layout of AMRScan 2. (A) The Nextflow workflow calls each assembly with AMRFinderPlus through the caller wrapper, records provenance and converts the report into a harmonised table. Shaded boxes are files written by the software. (B) The AST tool archives exports from NCBI Pathogen Detection and builds isolate and endpoint tables offline. The two are joined by assembly accession in the analysis that uses them.

Each sample is called in its own task by the AMRFINDERPLUS process and converted by the HARMONIZE_AMRFINDER process. Both processes run the same Python script (`bin/amrscan.py`) inside one container image that provides AMRFinderPlus 4.2.7 and Python 3.12. Docker, Apptainer or Singularity [12] and Conda (through Bioconda [13]) profiles are built from one pinned specification, and a profile for the PBS Pro scheduler of the NCI Gadi system is included. Tasks killed for memory or walltime are retried twice with more resources; any other failure stops the run.

The AMRFinderPlus database is fetched once by a separate script, which downloads a named release, indexes it and records the hash of every file. The workflow never downloads or updates a database, so all samples in a run are called against the same files. The organism option enables AMRFinderPlus's taxon-specific point-mutation screening for one taxon per run; without it the result is not a mutation screen.

## Caller wrapper and provenance

The caller wrapper checks the input against its expected SHA-256 when one is given, runs AMRFinderPlus with the requested settings, and writes three files per sample: the unmodified report, the caller's log, and a provenance record in JSON. The record holds the exact argument vector, the version reported by the executable, the SHA-256 of the input, the source path and per-file hashes of the database together with a single database fingerprint (the SHA-256 of the sorted map of relative file names and file hashes), the hash of the report and the completion status. The fingerprint identifies the database contents rather than its release name, so two copies of the same release that differ in any file are told apart.

## Harmonised output

The conversion step reads the report only if its hash matches the provenance record, and writes one row per element reported by the caller to a tab-separated table with a fixed schema: sample, element symbol, allele where the caller reports one, class and subclass, identity, coverage, coordinates and strand, method, element type and scope, the caller version, the database fingerprint and the thresholds used. Every original column is also kept in a JSON field, including columns that a later release of AMRFinderPlus might add. Rows are not deduplicated, ranked or interpreted: the schema has a drug column, but this adapter sets it to NA, and no resistant or susceptible call is made. A sample with no reported element gives a table with a header only, which records a completed call rather than susceptibility.

## Susceptibility data from NCBI Pathogen Detection

A separate command-line tool, `bin/ast_dataset.py`, archives AST and isolate tables from the NCBI Pathogen Detection table exporter (Fig. 1B). The `download` command takes a query, or a list of isolate accessions retrieved in batches, and checks each batch against an independent count request; the URL, retrieval time, row count and SHA-256 of every export are written to a snapshot manifest, and a failed request leaves no snapshot. The `build` and `validate` commands work offline. They keep AST rows at the level of the individual observation, including repeated tests, conflicting results and non-binary categories, collapse observations into isolate and antimicrobial endpoints with explicit status, and record BioSample and assembly accessions for linking to genomes. `validate` rebuilds the tables from the snapshot and compares them with an existing build byte for byte. Assemblies can then be fetched by accession with NCBI Datasets [14]; a helper script writes a samplesheet with the hash of each downloaded file.

AMRScan keeps the determinant calls and the AST data apart. Mapping determinants to antimicrobials, applying breakpoints and comparing genotype with phenotype are left to the analysis that uses them.

## Testing and documentation

The Python code has offline unit tests run in continuous integration with a 90% branch-coverage floor; parser tests use stored AMRFinderPlus reports and fail on missing or duplicated headers, malformed rows, invalid coordinates and changed report hashes. nf-test cases run each module and the whole workflow in the pinned container, and the `test` profile runs end to end on a positive control (*bla*$_{TEM-1}$ from NCBI reference sequence NG_050145.1) against a cached database release. A workshop for new users, notes on running on Gadi and decision records for design choices are published as a documentation site.

# Use on public *E. coli* genomes

## A 10,521-genome cohort on a workstation

AMRScan was used as the data layer of a study of genomic AST prediction in *E. coli* (manuscript in preparation). The AST tool archived exports for 10,521 NCBI Pathogen Detection isolates with AST results and a GenBank assembly, retrieved by exact accession in 106 batches of up to 100 isolates. Every batch matched its count request, the batches did not overlap and together they covered all isolates. The snapshot holds 162,701 AST observations, which collapse to 162,629 isolate and antimicrobial endpoints; all observations matched their isolate's BioSample and organism, and no BioSample or assembly mapped to two isolates.

The assemblies were called with the caller wrapper, driven by a batch script outside Nextflow, with AMRFinderPlus 4.2.7 (osx-arm64 Conda build), database release 2026-08-07.1 and the *Escherichia* organism setting, on an Apple M4 workstation with 16 GB of memory, eight workers and one thread per call. All 10,521 calls succeeded in 11.4 h of active wall time, at about 13 to 18 assemblies per minute. The run was interrupted several times; on resumption, outputs of 22 incomplete calls were set aside and called again, and 4,362 completed calls were reused only after their provenance records, report hashes and harmonised tables had been checked again. No provenance mismatch or failed conversion was recorded.

## 1,000 genomes through the workflow on an HPC system

To test the workflow at scale, a random 1,000 of these assemblies (5.1 Gb of sequence) were downloaded again on the NCI Gadi system, checked against the hashes recorded at the first download and run through the workflow with the Gadi profile, the same database release and the same settings, with two CPUs per task (Table 1). All 2,000 tasks succeeded without a retry. Each AMRFinderPlus task took a median of 24 s and peaked at a median of 268 MB of memory; with one scheduler job per assembly, tasks spent longer waiting in the queue than running.

Table 1. Run of 1,000 *E. coli* assemblies through the workflow on NCI Gadi (28 October 2026).

| Item | Value |
|---|---|
| Software | AMRFinderPlus 4.2.7 and Python 3.12 in one container image (SingularityCE 3.11.3); |

| Item | Value |
|---|---|
| | Nextflow 25.04.6 |
| Database | AMRFinderPlus release 2026-08-07.1, downloaded and indexed on the system |
| Tasks | 1,000 AMRFinderPlus tasks (one scheduler job each) and 1,000 conversions; no retries or failures |
| Wall time | 1 h 10 min |
| AMRFinderPlus task time | median 24 s; 95th percentile 37 s; maximum 5 min 35 s |
| Peak memory per task | median 268 MB; maximum 304 MB |
| Queue wait per task | median 4 min 10 s |
| Service units | 20.7 for tasks and 4.7 for the head job |
| Output | 9,441 harmonised rows; 187 assemblies with no reported element |
| Agreement with the workstation calls | all 1,000 reports byte-identical |

The AMRFinderPlus reports were byte-identical to those of the workstation run for all 1,000 assemblies, although one run used a native Conda build on Apple silicon and the other a Linux container. The two database fingerprints differed. The release directory on the NCBI site held five files that had not been there at the first download, and the BLAST and HMMER index files differed between the two indexing runs; every file present in both copies had the same hash. The fingerprint therefore separated two copies of the same named release, and the identical reports showed that the difference did not change a call. Without per-file hashes this difference would not have been visible.

## Relation to other tools

AMRScan 2 does less than the general workflows and is not a replacement for them (Table 2). It runs one detector on assemblies, where nf-core/funcscan runs several detectors and Bactopia covers a much wider range of analyses [5]. It does not interpret determinants as phenotypes, as abritAMR [3] and staramr [4] do, and its harmonised table is specific to AMRFinderPlus, whereas hAMRonization covers many detectors [6]. What it adds is a record, for every sample, that ties the reported elements to the exact input, software and database files, a check that refuses a report changed after the call, and an archive of the public AST data that the calls are later compared with. These features matter mainly when calls are used as data in genotype–phenotype studies or benchmarks of AST prediction [15], where an undetected change in input, database or phenotype export would change the results.

Table 2. Scope of AMRScan 2 and related tools, as described in their publications and documentation.

| Tool | Scope | AMR detection | Interpretation of determinants |
|---|---|---|---|
| AMRScan 2 | Determinant calls with per-sample provenance; archive of NCBI Pathogen Detection AST data | AMRFinderPlus | None; calls and phenotypes kept apart |
| abritAMR [3] | AMRFinderPlus for accredited clinical and public health use | AMRFinderPlus, with species-specific mutations | Drug classes; *Salmonella* phenotype prediction |
| staramr [4] | Antibiogram prediction from assemblies | ResFinder, PointFinder, PlasmidFinder | Gene–drug key giving a predicted antibiogram |
| Bactopia [5] | Complete analysis of bacterial genomes | AMRFinderPlus among many tools | Not assessed here |
| nf-core/funcscan | Functional screening of contigs | ABRicate, AMRFinderPlus, RGI, DeepARG, fARGene | Outputs combined with hAMRonization |
| hAMRonization [6] | Common specification for detector outputs | Parsers for many detectors | Common drug-class fields |

# Limitations and planned work

The workflow accepts only assembled nucleotide sequences; reads, combined protein and annotation searches and per-sample organism settings are not supported, and the organism option applies to the whole run. Only AMRFinderPlus is wrapped; adapters for RGI with CARD and for ResFinder are planned and will have to fill the same schema and keep their own reports and provenance. Hashing every database file for every sample adds input and output on large cohorts. The AST tool reads whole tables into memory, has no automatic retry or resume and does not reconcile different releases of the same records; adjudication of AST methods and breakpoints is left to the user. Parts of the workflow are being prepared for contribution to nf-core.

## Changes from version 1

Version 1 of this preprint described AMRScan 1: a BLAST search against the CARD protein homolog model, run from an R script or a single Nextflow file, with a top-hit summary as output. Version 2 describes a rewrite. The detector is AMRFinderPlus, the workflow is modular, software comes from one pinned specification, every run is recorded and checked, the output keeps the full caller report, and a tool for archiving NCBI Pathogen Detection AST data has been added. The repository has also moved, to https://github.com/taowis/amrscan.

## Availability

AMRScan is available at https://github.com/taowis/amrscan under the MIT licence, with documentation at https://taowis.github.io/amrscan/. The version described here is release v2.1.0, archived at https://doi.org/10.5281/zenodo.23825954. The scripts, inputs and run records of the Gadi run are in `examples/gadi-scale/`, and the hashes of the 10,521 assemblies are in `benchmark/cohorts/ecoli_research_v1/genome_manifest.tsv`. Version 1 is in `legacy/`.

## Funding

This work received no specific grant from any funding agency in the public, commercial, or not-for-profit sectors.

## Acknowledgements

Computing on Gadi was provided by the National Computational Infrastructure, Australia.

## Conflicts of interest

The author declares that there are no conflicts of interest.